# Greek Astronomy PhDs: The last 200 years…

Vassilis Charmandaris
University of Crete & National Observatory of Athens

Abstract

We have recently compiled a database with all doctoral dissertations (PhDs) completed in modern Greece (1837-2014), in the general area of astronomy and astrophysics, as well as in space and ionospheric physics. A preliminary statistical analysis of the data is presented along with a discussion of the general trends observed.

Introduction

One of the principal metrics of the performance of a country in a given academic discipline is the research output, as documented by the refereed publications produced by the researchers in this country, and the impact these publications have in the international community, as measured by the citations they have received. Another related index is the number and quality of doctoral dissertations that have been completed in the academic institutions of the country and the subsequent academic progress of the PhD holders in the field.

In this report I present a first analysis related to the last topic, based on a recently compiled database of all doctoral dissertations (PhDs) completed in Greek academic institutions in the general area of "Astronomy". The meaning of the term "Astronomy" here is rather broad, since we include not only the fields of classical observational astronomy, astrophysics, and dynamical astronomy, but also the areas of modern space physics and ionospheric physics. The period we cover effectively starts in 1886, when Prof. Demetrios Eginitis - the fifth director of the National Observatory of Athens - obtained his PhD degree from the University of Athens and it ends in 2014.

One should note that even though the University of Athens was founded in 1837, there are no records of PhDs awarded in the field of observational or dynamical astronomy prior to the arrival of Eginitis. The first three Professors of Astronomy had no PhD students. Two of them were educated abroad. George Bouris, Professor during 1837-1855, was educated in Vienna and Demetrios Kokkides, Professor during 1877-1896, obtained his PhD in Berlin, in 1862. Ioannis G. Papadakis, Professor during 1850-1876, was actually a mathematician, with no published record related to astronomy. It remains unclear who the supervisor of Eginitis was, since at the time no extended thesis was required for someone to obtain a PhD degree. It seems though that Eginitis (PhD Univ. of Athens, 1886) was greatly influenced[1] by Cyparissos Stephanos, a Professor of mathematics at the Univ. of Athens (1884-1917) who had obtained his PhD at the University of Paris in 1884.

Prof. Eginitis, who held the Chair of Astronomy at the University of Athens for 38 years (1896-1934) and served as the Director of the National Observatory of Athens for 44 years (1890-1934), supervised one PhD student; Stavros Plakidis (PhD Univ. of Athens, 1931)[2]. When Prof. Eginitis passed away[3], Plakidis succeeded him at the Chair of Astronomy at the University of Athens, held it for another 30 years (1935-1965) and supervised 8 PhDs. This is the principal "root" of most Greek astronomy PhDs today: a total of 75 PhDs are decedents of Prof. Plakidis.

Finally, it should be stressed that even though the assistance of several colleagues was essential for the compilation of data, the analysis and opinions expressed in the article are of the author and may be subjective in nature.

---

[1] See "In memoriam of Demetris Eginitis", by C. Carapiperis et al. (1975)
[2] Note that A.S. Eddington is acknowledged by S. Plakidis as his co-supervisor, since he suggested the thesis topic and they published a paper together (Eddigton & Plakidis 1929)
[3] See "Demetrios Eginitis: Restorer of the Athens Observatory", by Theodossiou et al. (2007)

## The Data[4]

To collect the data we relied on a number of sources including the records of Greek Universities, previous reports with biographies of Greek astronomers (Laskarides 1992, 2009), feedback from senior colleagues, as well as the dissertation archive of the National Documentation Center. The final output was 245 PhDs from 7 academic institutions[5]. With the exception of just four nationals from Egypt, Poland, Italy and Bulgaria, all PhD recipients were Greek. For each record we verified the details of the individual awarded the degree, her/his current position, the awarding institution, the supervisor name, as well as the thesis title. For 129 of them (53%) the full manuscript of the thesis is also available online. Even though we are complete regarding the dissertations performed at the Univ. of Crete, Univ. of Ioannina, and Univ. of Thrace, it is hard to accurately estimate the completeness of the data, in particular for the oldest University of Athens and also the University of Thessaloniki. However, a "guestimate" of the author is that the uncertainties are less than 3%, and do not affect the global statistics and trends discussed in this study. Any missing data can easily be filled in once provided.

The distribution of the PhD degrees awarded by academic institution is presented in the following Figure 1.

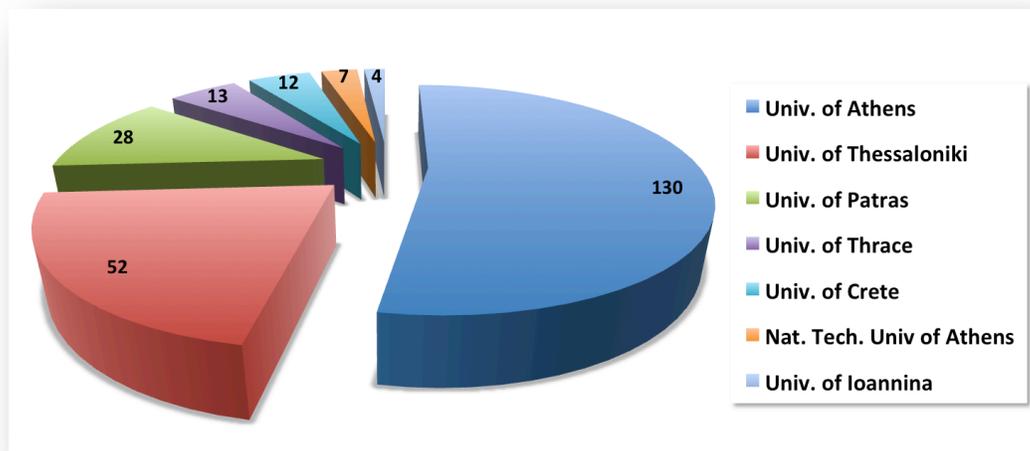

**Figure 1**. A pie chart of the 245 dissertations performed at the 7 major Universities in Greece, from 1886 through 2014.

## Analysis

As it is common in most countries, only Universities can award PhD degrees in Greece. These were typically performed at the various Departments of Physics and Mathematics. Up until the late 1970s though, there were Chairs of Geodetic Astronomy both at the Univ. of Thessaloniki and the National Technical University of Athens (NTUA). As a consequence, 7 dissertations in dynamical astronomy and/or geodesy were completed at NTUA. Moreover, over the last 30 years, the Electrical and Computer Engineering Department at the University of Thrace hosts an active Space Physics group and has produced 13 PhDs in this area.

Due to the restructuring of the Greek academic system in 1982, most research in astrophysics progressively moved out from the engineering schools. Currently, it is performed mainly at the Departments of Physics across the country, as well as at the research institutes of the National Observatory of Athens and the Academy of Athens.

---

[4] "Without data you're just another person with an opinion", quote by Dr. E. Deming.
[5] The database with all ancillary information is available at http://www.helas.gr/documents.php

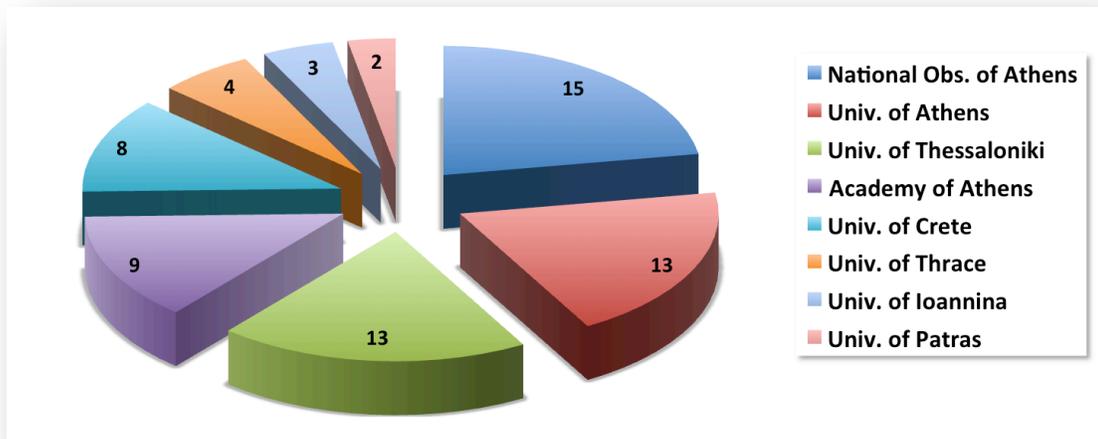

**Figure 2**. A pie chart of the 67 tenure or tenure track faculty and researchers[6] in the field of astrophysics, dynamical astronomy, and space physics at the various Universities and Research Institutes in Greece. Of those positions 11 are current staffed by females, 16% of the total.

It is of particular interest to examine the employment of the past PhD recipients, as this provides some information, not only about the quality of the research performed during their thesis, but also about the evolution of the job market in Greece.

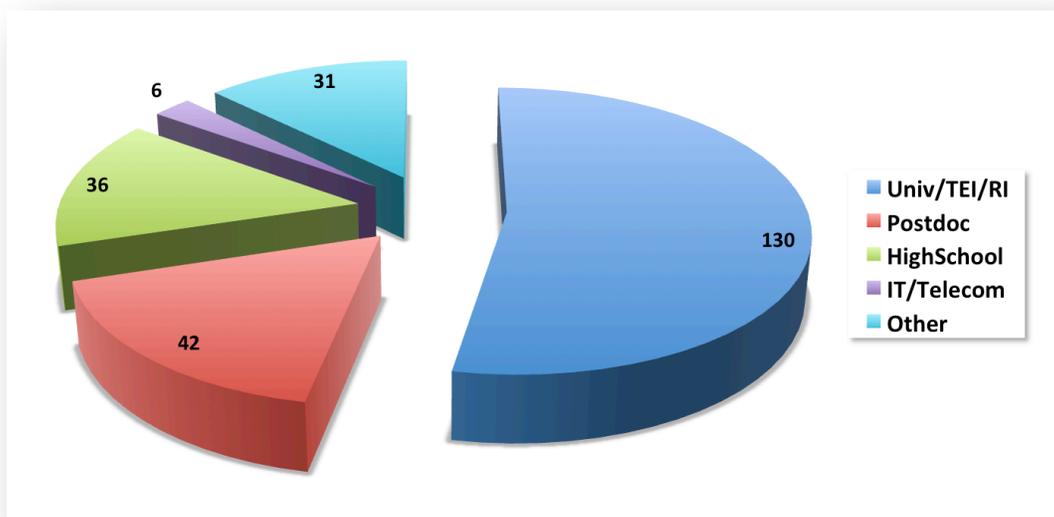

**Figure 3**. A pie chart of the current (July 2015) position of the 245 PhD recipients. For individuals who retired or passed away, the last position is indicated. Note that 27% are in postdoc/soft-money positions, while 15% have moved to secondary education.

We decided to divide the current jobs of the PhD recipients in 5 categories. The first includes permanent positions for which the degree is a requirement, such as a faculty at a University and Technological Educational Institute (TEI), or a researcher and research support specialist at Research Institutes (RI). The second is a postdoctoral researcher position and this includes mostly junior researchers still in soft-money positions, a few years after their PhD. The third includes individuals that left academia and teach in high schools. The fourth is for people who moved to the private sector in fields such as Information Technology (IT) support and telecommunications. We also have a category for everything else.  Obviously, this information is subject to change, in particular for those in non-permanent positions (i.e. Postdoc or Other categories). For individuals who retired or passed away, we indicate their last position.

---

[6] Among the 67 individuals we also include 6 who are currently on extended leave of absence from their home institute while working abroad, and it is unclear whether they will return back to Greece.

The job distribution, as of July 2015, is presented in Figure 3. We note that more than half (53%) of the individuals have obtained a permanent position, for which a PhD was a requirement. One should also note that among them there are 7 individuals who obtained a permanent or long-term position outside Greece. At first glance, this would indicate that the current prospects for an Astronomy PhD awarded in Greece are rather good. However, as we can see in Figure 4, this may not necessarily be the case…

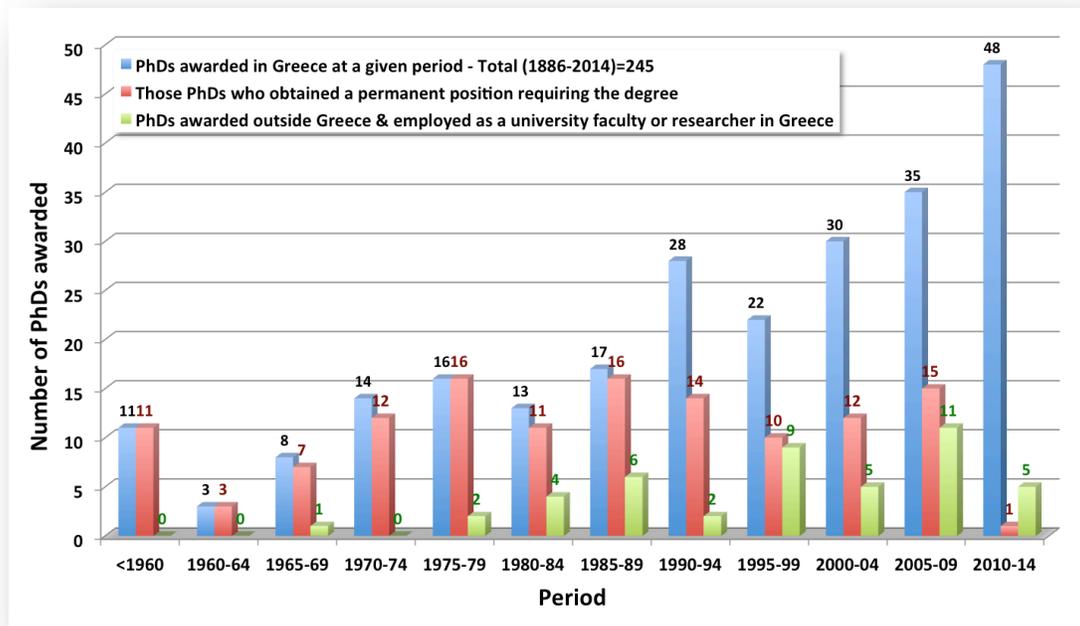

**Figure 4**. Histogram of the "Astronomy" PhDs awarded in Greece binned in 5-year periods. The degrees awarded prior to 1960 are grouped together (see text).

In Figure 4 we present in blue the number of degrees awarded, binned in 5-year periods. In red we indicate whether the individuals who obtained their degree within the given timespan eventually obtained a permanent position that required a PhD, either in Greece or abroad. This effectively means that the individual obtained a job either as a faculty at a University/TEI, or that she/he found a tenure track (or is currently in a long term) position as a researcher or research support personnel at a Research Institute. Note that it is not implied that this permanent position was obtained within the 5-year period the PhD degree was awarded.

In the same histogram we indicate in green the number of tenure track positions (faculty positions at Universities or full time researchers at Research Institutes) in Greece, which were filled by scientists who obtained their PhD outside Greece. One should note that these are individuals who had no prior employment connection with Universities in Greece. We have to stress this, since in the 1960's and 70's there were several individuals who at that time had obtained semi-permanent positions (the so called "Assistants" without a PhD), either at Universities or at Research Institutes. Some of them left Greece, obtained their PhD abroad and returned to their previous position, which later became permanent. Since this established prior connection existed and likely gave them an advantage in obtaining tenure, we decided not to consider this group as truly "outsiders".

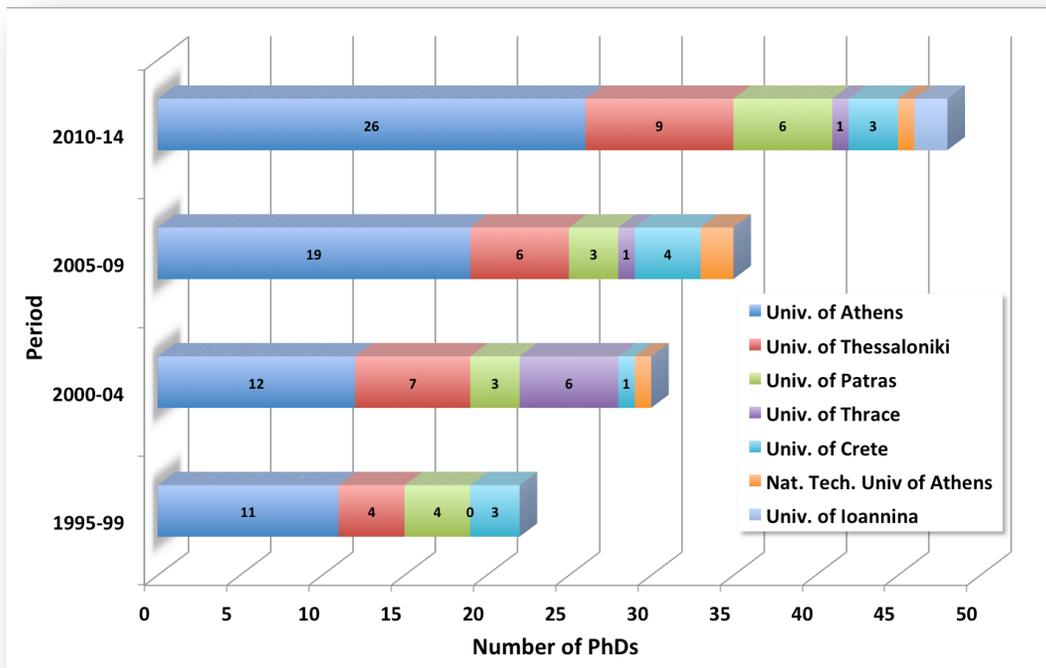

**Figure 5**. Histogram of the "Astronomy" PhDs awarded, since 1995, by the various Universities in Greece, binned in 5-year periods.

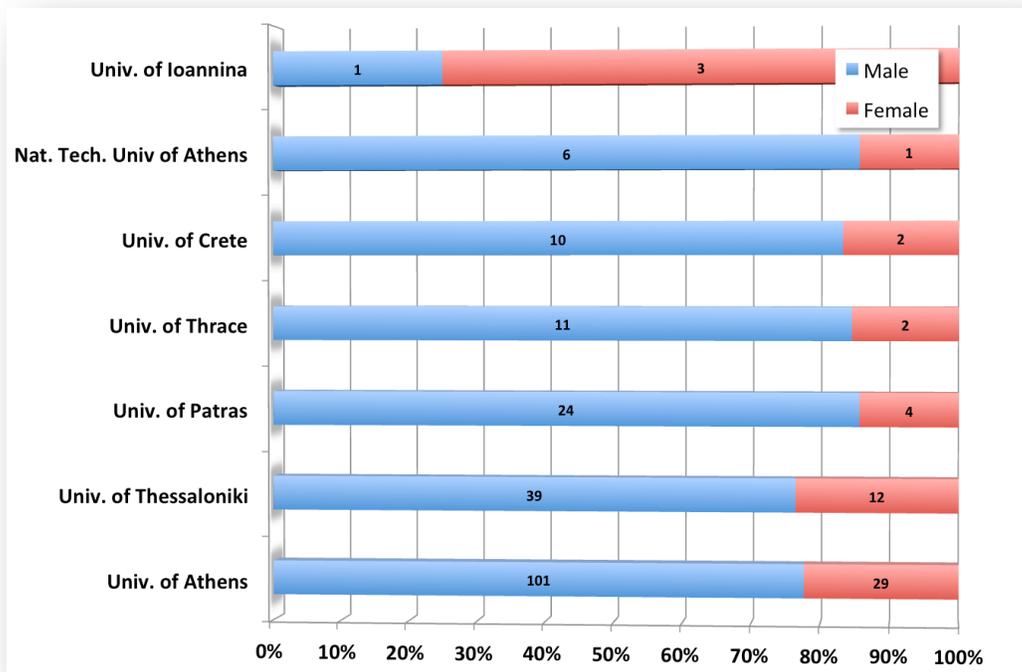

**Figure 6**. Histogram of the distribution of the 245 PhDs awarded by Greek academic institutions, broken by gender. There are 53 PhDs awarded to women, 22% of the total.

It is instructive to briefly discuss a few conclusions one may draw from Figures 4, 5 and 6 in more detail.

- ❏ The number of PhDs has nearly tripled over the last 30 years and doubled over the last 20. This is likely related to the increase in the number of University faculty in mid-80's, as well as the available funding from research grants supporting graduate studies. Moreover, the first graduate programs leading to a Masters degree were formally legislated in 1992, attracting more students who instead of leaving abroad, stayed in Greece with some of them continuing to a PhD.
- ❏ It is obvious that up until the end of the 1980s nearly every one of the 81 PhD recipients eventually obtained a permanent position in the academia. This was probably due to the expansion of the academic system, the small number of PhD students, and the limited competition from individuals with PhDs awarded abroad.
- ❏ Past 2010, with one exception (a foreigner who obtained a tenure track position in her country), no one has obtained a permanent position yet. This is expected though since the current stiff competition typically requires at least two postdocs (5-6 years research experience past the PhD) before someone is sufficiently competitive to obtain a tenure track position.
- ❏ During the period between ~1990 and 2005 a third of the PhDs obtained a permanent academic job. Some of these hires were due to favorable legislation (i.e. law 1268 in 1982, as well as those by V. Papandreou and P. Pavlopoulos later), which gave the possibility to soft money/postdoctoral positions to be converted to permanent faculty/researchers with rather relaxed academic criteria. It has been argued that several of the problems that plagued the Greek academic system since then, are directly related to these laws.
- ❏ After the 1980s the number of PhDs from abroad who obtained a permanent position in Greece steadily increases. In fact, if one focuses only on the 67 faculty and researchers currently in tenure-track positions, 49% of them are Greeks who have obtained their PhD outside Greece. The fraction is even higher (63%) if we restrict on the hires over the past 10 years. This "extrovert nature" in hires is very positive, in particular since several of the individuals with Greek PhDs in a permanent position, have also spent substantial amount of time outside Greece before returning back, bringing with them international practices in doing high quality research as well as and on the way of thinking.
- ❏ Regarding the gender issue, a rather pertinent topic in higher education and in particularly in the areas of Science and Engineering, the fraction of "Astronomy" PhDs that has been awarded to females in Greece is 22%. This fraction goes up to 27% if we restrict our analysis in the past 20 years. Furthermore, there are 11 females among the 67 individuals with tenure positions, that is 16% of the total. In 2006, when a similar analysis had been performed, the fraction was 13% (Charmandaris 2006). These numbers are typical of other western countries[7]. It should be noted thought that so far there has never been a female full professor in Astronomy in Greece. There has been only one female astronomer at the top level of "Researcher A", who has retired in 2008.

## Perspectives

Keeping in mind that "it's hard to predict things, especially if they pertain to the future"[8], can one speculate what the future holds for employment in permanent "Astronomy" related academic positions in Greece? We may use Figure 7 as a guide on this endeavor.

The legend of Figure 7 indicates the average age and its standard deviation of the 67 permanent faculty/researchers in the eight Universities and Research Institutes listed in Figure 2. The astronomy sections at the Universities of Athens and Thessaloniki are fairly aged – albeit with a large dispersion - while the corresponding groups at the Academy of Athens and the University of Crete are the youngest. The groups at the University of Patras and Ioannina, comprising of 2 and 3 members respectively, are rather small for meaningful statistics to be drawn.

---

[7] For more information visit http://iauwomeninastronomy.org
[8] A fairly well known "deep thought" by Yogi Berra.

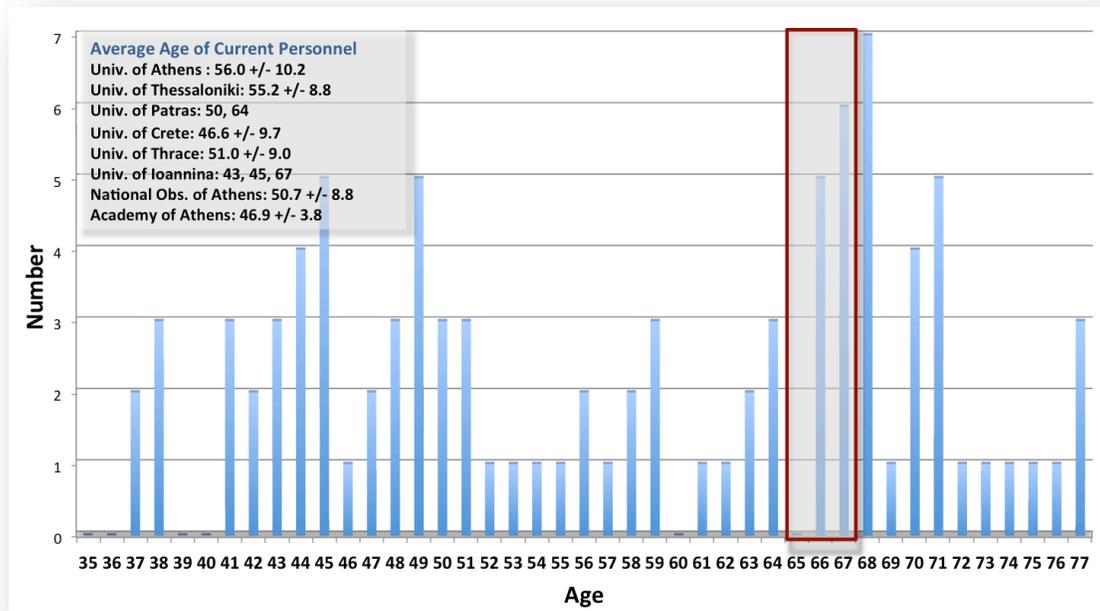

**Figure 7**. Histogram of the age distribution of all individuals currently on (or recently retired from) an "Astronomy" tenure or tenure-track position in Universities or Research Institutes. The grey shaded red box marks the ages when retirement is compulsory (65 in Research Institutes and 67 in Universities). Note that at the time of this report (July 2015) some individuals have already retired, even though they have not reached this age.

**Table 1:** "Astronomy" hires and retirements (in parenthesis) from Greek Institutions.

| Institution | 2005-2009 | 2010-2014 | P | 2015-16 | 2016-2019 |
|---|---|---|---|---|---|
| Univ. of Athens | 1 (4) | 3 (11) | 13 | (2) | (1) |
| Univ. of Thessaloniki | 1 (1) | 3 (2) | 13 | (4) | (1) |
| Univ. of Crete | 2 (2) | 2 (1) | 8 | (2) | (0) |
| Univ. of Thrace | 0 (0) | 1 (1) | 4 | (0) | (0) |
| Univ. of Ioannina | 2 (1) | 0 (1) | 3 | (1) | (0) |
| Univ. of Patras | 0 (0) | 0 (2) | 2 | (0) | (1) |
| Nat. Obs. of Athens | 4 (1) | 0 (2) | 15 | (1) | (2) |
| Academy of Athens | 4 (3) | 1 (0) | 9 | (0) | (0) |
| **Total:** | **14 (12)** | **10 (20)** | **67** | **(10)** | **(5)** |

As we display in Table 1, over the period 2005-2014, a total of 32 individuals retired from permanent faculty/researcher positions. The number of new hires over the same period was 24 (just over 2 new positions per year). Under the column entitled "P" we indicate the current (July 2015) tenure and tenure track personnel in each Institution. Another 10 persons will have to retire by August 2016. Given the delays associated with the bureaucracy of new hires, it is safe to assume that no one will be hired over the same period. This will create an overall deficit of 18 permanent positions, or 27% compared to 10 years ago. The situation will get worse by the end of 2019, when another 5 permanent researchers will reach the retirement age.

Even if the Universities and Research Institutes will likely not reach the size they had in the late 1980s, some of the void in personnel will have to be filled in order for them to be able to function effectively. Assuming that the current economic crisis is resolved soon, the possibilities to obtain a permanent academic position in "Astronomy" in Greece will increase over the next 5 to 10 years. It is unclear though how attractive this option will be for high quality individuals with career options abroad, unless the reduction in salaries and support to research and infrastructures that took place over the past 5 years is also reversed. Given the events of early July 2015, both prospects seem rather unlikely to the author.

The future of available soft-money positions is brighter. Over the past 15-20 years there are nearly constant funding possibilities mostly from the European Commission and the European Space Agency, as well as some national structural funds. This is not expected to change in the near future. Moreover, the quality of the permanent researchers and university faculty in Greece has been steadily increasing and many of them have strong international connections with leading academic institutions abroad. This should facilitate their ability to attract funding and to support personnel and research infrastructures.

## Acknowledgments

I greatly appreciate the support of all colleagues who provided the information to complete and validate the database of the Greek "Astronomy" PhDs. I wish to thank in particular Prof. H. Varvoglis and J.H. Seiradakis, who provided substantial help mostly regarding the records from the Univ. of Thessaloniki, as well as Prof. T. Kalvouridis for those from the Univ. of Patras and NTUA. The help of Ms. S. Zarbouti, the Secretary of the Section of Astronomy, Astrophysics, and Mechanics at the Univ. of Athens was also simply invaluable. Finally, I would also like to acknowledge the help of Prof. N. Kylafis with whom I had fruitful discussions on various topics related to this article. Nick also took the time to carefully read the article on a short notice, leading to a number of clarifications as well as corrections of various typos. I hope that the database will not only preserve a part of the history of modern Greek astronomy, but it will also be beneficial for other more thorough studies in the future.

This report was finalized on June 30, 2015. It has last been updated on July 7, 2015.